\documentclass[twocolumn,10pt]{article}

\usepackage{graphicx}

\usepackage{amsmath,amssymb} 
\usepackage{gensymb}
\usepackage{mathtools}
\usepackage{mathrsfs}
\usepackage{lipsum}
\usepackage{float}
\usepackage{pgfplots}
\usepackage{graphicx,amsmath,upgreek,bm}
\usepackage{subcaption}
\usepackage{siunitx}
\usepackage{makecell}
\usepackage{booktabs}
\pgfplotsset{compat=newest}

\pgfplotsset{plot coordinates/math parser=false}

\graphicspath{{./figures/}{../figures/}{.}}
\usepackage[margin=1in]{geometry}
\usepackage{multirow}
\usepackage{graphicx}
\usepackage{dcolumn}
\usepackage{bm}

\usepackage{nameref}
\usepackage{lipsum}
\usepackage{multirow}
\usepackage{graphicx}
\usepackage{dcolumn}
\usepackage{bm}
\usepackage{newtxtext}
\usepackage[utf8]{inputenc}
\usepackage[T1]{fontenc}
\usepackage{mathtools}
\usepackage{mathrsfs}

\usepackage{mathtools}
\usepackage{ mathrsfs }

\usepackage{amsmath, amssymb}
\usepackage{authblk}

\title{Effects of Correlated Noise on the Performance of Persistence Based Dynamic State Detection Methods}

\author[1]{Joshua R.~Tempelman}
\author[2]{Audun D.~Myers}
\author[3]{Jeffrey T.~Scruggs}
\author[4]{Firas A.~Khasawneh}

\affil[1,2,4]{Dept. of Mechanical Engineering, Michigan State University.}
\affil[3]{Dept. of Civil \& Environmental Engineering, University of Michigan}
\affil[4]{Email: khasawn3@egr.msu.edu}

\begin{document}

\maketitle    

\begin{abstract}
The ability to characterize the state of dynamic systems has been a pertinent task in the time series analysis community. Traditional measures such as Lyapunov exponents are often times difficult to recover from noisy data, especially if the dimensionality of the system is not known. More recent binary and network based testing methods have delivered promising results for unknown deterministic systems, however noise injected into a periodic signal leads to  false positives. Recently, we showed the advantage of using persistent homology as a tool for achieving dynamic state detection for systems with no known model and showed its robustness to white Gaussian noise. In this work, we explore the robustness of the persistence based methods to the influence of colored noise and show that colored noise processes of the form $1/f^{\alpha}$ lead to false positive diagnostic at lower signal to noise ratios for $\alpha<0$.
\end{abstract}
\section{Introduction}

The distinction between regular and chaotic dynamics has been a thoroughly researched topic in the fields of applied mathematics and engineering~\cite{Bagtzoglou2007,Zhang2003,Avanco2016}, signal analysis~\cite{Melosik2016,Leung1993}, and biological systems~\cite{olsen1985}. Often times, the detection of chaos versus periodicity is desired to make inferences about the predictability and stability of the system.  However, detecting chaos in time series data is not a straightforward procedure, and many studies have been devoted to this~\cite{Gottwald2004,Wernecke2017,sandri1996numerical,Benettin1980,Tempelman2019,Myers2019}. While many tools have proven to be effective for analyzing toy models (e.g. the Lorenz and Rossler models), the effects of noise on chaos detection methods are often overlooked. 

Noise is an inherent factor in all natural and engineered systems. Thus, the effects of noise on deterministic systems must be explored for all proposed signal analysis schemes. 
The effects of noise on determining the dynamics have been explored through a variety of works pertaining to Lyuponov exponents, the 0--1 test, embedded networks~\cite{Myers2019}, and TDA based methods~\cite{Liu2005,Wolf1985}. The effect of noise on the more recent 0--1 test for chaos was explored in~\cite{Gottwald2005,Gottwald2009,Tempelman2019} and it was shown in~\cite{Gottwald2009,Tempelman2019} that this method can be made very noise robust. 
However, these studies are all restricted to the effects of uncorrelated Gaussian noise models. 

While noise in dynamic systems often present as Gaussian, it is not uncommon to find correlated color noises as well. Correlated noise, often called colored noise, is an alternative model to the broadband Gaussian noise. 
Colored noises can appear in both mechanical and electrical systems~\cite{Schueller2006,Ding2013}. A common application of colored noise phenomenon is as a model for sensors and actuators~\cite{kasdin1995discrete}. Thus, the effects of correlated noise processes on the performance of detection methods is of interest to the engineering design and analysis community. 

This paper addresses the colored noise question by exploring various effects that colored noises have on the recently developed chaos detection methods given in~\cite{Gottwald2016,Tempelman2019,Myers2019}. 
Simulations of the Lorenz and Rossler models are used to generated time series which display both chaotic and periodic dynamics. 
Various noise colors are synthetically generated and added to the simulation data at a variety of intensity levels. 
Some chaos diagnostic tools are then applied to these data, and the effects of the various noise colors are reported for each system. 

The organization of this paper is as follows. We begin by introducing the 0--1 test for chaos and the concept of ordinal partition networks in section~\ref{sec:Testing_for_chaos}. An introduction to persistent homology is given in section~~\ref{sec:sublevel}. Section~\ref{sec: detecting_chaos_w_pers} outlines persistence based methods for detecting chaos in dynamics systems with an extension to the 0--1 test given in section~\ref{sec:PS Test} and an extension to ordinal partition theory given in section~\ref{sec: persistence_on_networks}. 
 The descriptions of the noise models explored in this work is given in section~\ref{sec: Noise}. The results of this study and summarizing conclusions are given in sections~\ref{sec: Results} and~\ref{sec: Conclusion} respectively. 


\section{Testing for Chaos} \label{sec:Testing_for_chaos}
While tracking the sign of the maximal Lyapunov exponent has historically been the de-facto standard for quantifying chaotic dynamics, this tool can often times lead to ambiguity and is difficult to apply to systems with no known model~\cite{sandri1996numerical, Benettin1980} as well as for systems with noise~\cite{Liu2005}. Recently, two new methods have been developed to detect chaos. The first method is based on Gottwald and Melbourne 0--1 test and is introduced in Section~\ref{sec:original_01_test}. Their work produced a method for providing a definitive 0--1 diagnostic for chaos in deterministic systems, which works conveniently for a single time series~\cite{Gottwald2004,Gottwald2005,Gottwald2008,Gottwald2016}. The second method, introduced in Section~\ref{sec:orginal_part_network}, is based on an analysis of ordinal partition networks developed by McCullough et al.~\cite{McCullough2015}.

\subsection{The original 0--1 test} \label{sec:original_01_test}
The reader is directed to~\cite{Gottwald2016} for a full explanation for the original 0--1 (regression) test and well as the 0--1 (correlation) test. In this paper, the correlation test is used unless otherwise stated. 

Given some time series from a deterministic system $\phi(t)$, the $p$-$q$ planar data set is constructed according to 
\begin{equation}
p(n) = \sum_{j=1}^{n}\phi(j)\cos{jc},  \hspace{15pt}  {\mathrm{ and }}   \hspace{15pt} q(n) = \sum_{j=1}^{n}\phi(j)\sin{jc},
\label{eq:pq}
\end{equation}
where the random variable $c$ is selected from the uniform distribution $(0, \pi)$ and $n=\left(0,1,\dots,{N}\right)$. 
Drawing $N$ values of $c\in(0,\pi)$ leads the the construction of $N_c$ data sets in $(p,q)$ space. 
We use $N_c= 100$ and $N = 5000$.  The mean-square displacement $M_c(n)$ of these $p$-$q$ sets are then computed with
\begin{equation}
M_c(n) = \lim_{N \to \infty}\frac{1}{N}\sum_{j=1}^{N}\left[p_c(j+n)-p_c(j)\right]^2 + \left[q_c(j+n)-q_c(j)\right]^2.
\label{eq:MSD}
\end{equation}

	Note that in Eq.~\eqref{eq:MSD}, $n = \left(1,2\dots, n_{\rm cut}\right)$ where $n_{\rm cut} = N/10$. The 0--1 test is based on the asymptotic growth of $M_c(n)$ with respect to $n$, with has a linear logarithmic growth if the dynamics are chaotic~\cite{Gottwald2009}.  
	In~\cite{Gottwald2009} the modified mean-square displacement $D_c(n)$ is introduced which has better convergence characteristics.  
	Begin by computing the oscillatory terms
	\begin{equation}
	V_{\rm osc}(c,n) = \left(E\phi\right)^2\frac{1-\cos\left(nc\right)}{1-\cos\left(c\right)}.
	\end{equation}
	Now, the modified mean-square displacement is computed as $D_c(n) = M_c(n) - V_{\rm osc}.$ A 0--1 diagnostic can now be delivered using either the a correlation measure of by computing the linear regression of the log scaled modified mean-square displacements. The correlation test uses the standard definitions of covariance and variance,
	\begin{align}
	{\rm cov}\left(x,y\right) &= \frac{1}{q}\sum_{j=1}^{q}\left(x\left(j\right)-\bar{x}\right)\left(y\left(j\right) - \bar{y}\right),
	\hspace{10pt}{\rm and}\hspace{10pt}  \\
	{\rm var}\left(x\right) &= {\rm cov}\left(x,x\right), \nonumber
	\end{align}
	from which correlation $K_c$ can be computed via
	\begin{equation}
	K_c = {\rm corr}\left(\xi, \Delta\right) = \frac{{\rm cov}\left(\xi, \delta\right)}{\sqrt{ {\rm var}\left(\xi\right){\rm var}\left(\Delta\right)}}\in[-1,1],
	\label{eq:corr}
	\end{equation}
	where $\xi = \left(1,2,\dots,n_{\rm cut}\right)$ and  $\Delta = \left(D_c(1), D_c(2), \dots D_c\left(n_{\rm cut}\right)\right)$.
	The 0--1 correlation score is taken as the median of many $K_c$ where a score near 0 indicates periodicity and a score near 1 indicates chaos.

The alternative regression method is implemented with
\begin{equation}
K_c = \lim_{n\to\infty}\frac{\log{\tilde{D}_c(n)}}{\log{n}}, \ \ \ \tilde{D_c}(n) = D_c(n)-\min_{n=1,\dots,n_{\rm cut} }D_c(n).
\end{equation}
%
For finite data, the $K_c$ score is computed as the slope of the line which fits the $\log(M_c(n))$ versus $\log(n)$ with the least absolute deviation~\cite{Gottwald2009}.

\begin{figure}[h]
	\centering
	\includegraphics[width=.5\textwidth]{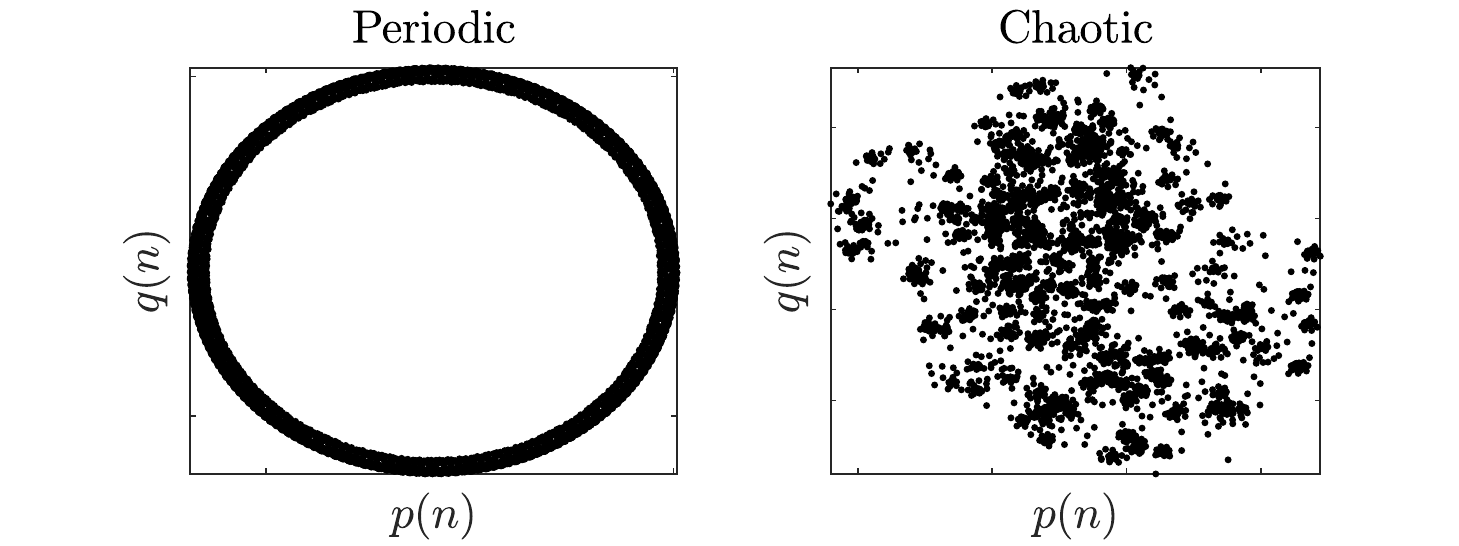}
	\caption{Examples of the $p$-$q$ projections of a (left) periodic and (right) chaotic time series.  }
	\label{fig:p-q projections}
\end{figure}

While the numeric computations of the 0--1 test deliver definitive results, it is noted in~\cite{Gottwald2004,Melosik2016} that the $p$-$q$ projections generated during the test are qualitatively different for periodic versus chaotic time series. 
The projections of $p$-$q$ data is typically regular and confined to a circular geometry for periodic time series, while chaotic time series lead to diffuse projections which emulate Brownian motion~\cite{Gottwald2004} (see  Fig.~\ref{fig:p-q projections}).

Following the work of Melosik and Marszalek~\cite{Melosik2016, Myers2019a}, the time series are first sub-sampled before being projected into $p$-$q$ space. A maximum significant frequency $f_{\rm max}$ approach is applied whereby the spectral content of the signal is used to determine the appropriate sampling frequency $f_s$ such that $2f_{\rm max}<f_s<4 f_{\rm max}$. We sub sample the data so that $f_s$ is three times the maximum significant frequency unless otherwise stated. 

\subsection{Ordinal Partition Networks} \label{sec:orginal_part_network}
An alternative method for detecting chaos is through the lens of networks. 
Networks can serve as a time series analysis tool by allowing for a unique interpretation and visualization of high-dimension dynamics through symbolic transitions that are captured in a graph. A network or graph $G = (V,E)$ is formed from a collection of vertices $V$ and edges $E$. For complex networks, the edges and vertices are derived from a time series, which is useful for capturing characteristics of peridoic in comparison to chaotic dynamics. There are several methods for embedding a time series into a networks including recurrence networks~\cite{Donner2010}, nearest neighbor networks~\cite{Khor2016}, and ordinal partition networks~\cite{McCullough2015}. 
These time series embedding to complex network tools generate graphs based on the underlying structure of the recunstructed state space (using Takens' embedding), which is critical for analyzing the dynamic state. In our previous work~\cite{Myers2019} we investigated both nearest neighbor and ordinal partition networks for dynamic state detection. However, we found that ordinal partition networks provided more consistent results and a faster computation time in comparison to nearest-neighbor networks. Therefore,  we will introduce and implement ordinal partition networks as a complexity measure to detect chaos. Ordinal partitions, or more commonly known as permutations, are a method of summarizing the time-ordered ranking of embedded vectors (through Takens' embedding) from a time series. Permutations were first popularized as a time series analysis tool through the complexity measure of permutation entropy~\cite{Bandt2002}. However, this method only analyzes the permutations through entropy as a statistical summary and does not account for the temporal order of the permutations. Because of this loss of information, ordinal partition networks were introduced, which capture the time-ordered sequence of permutations through a network.

\begin{figure}[h] 
    \centering
    \includegraphics[scale = 0.525]{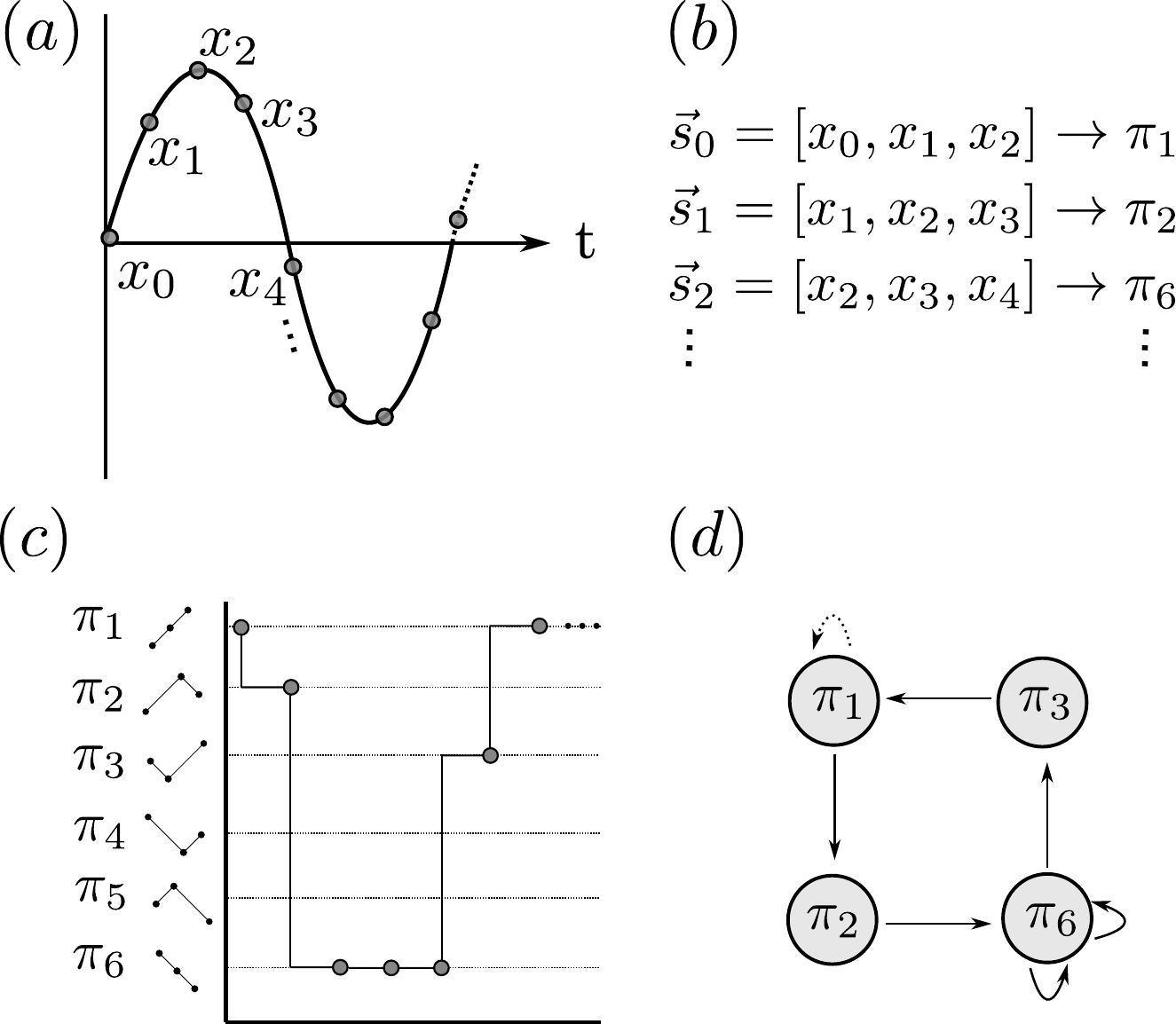}
    \caption{Example formation of an ordinal partition network in Fig.~(d) from a time series in Fig.~(a) using embedding dimension $n=3$ and delay $\tau=1$ to form embedding vectors from Takens' embedding as $\vec{s}$ in Fig.~(b).}
    \label{fig:timeseries_to_ordinal_partition_network}
\end{figure}

To demonstrate how to embed a time series into an ordinal partition networks we implement a simple example as shown Fig.~\ref{fig:timeseries_to_ordinal_partition_network}.
Using the time series in Fig.~\ref{fig:timeseries_to_ordinal_partition_network}-(a) as a collection of time-ordered data points ${\bf{x}} = [x_1, x_2, x_3, x_4, \ldots]$, we can embed the vectors into $\mathbb{R}^n$ with embedding delay $\tau$ using Takens' embedding theorem as $\vec{s}_i = [x_i, x_{i+\tau}, x_{i+2\tau},\ldots, x_{i+(n-1)\tau}]$. For this analysis, $n$ and $\tau$ are selected using multiple-scale permutation entropy as suggested in both~\cite{Myers2019a} and~\cite{Riedl2013}. We then embed the vectors using Takens' embedding $n=3$ and $\tau = 1$ as shown in Fig.~\ref{fig:timeseries_to_ordinal_partition_network}-(b). We can categorize the embedded vectors as one of $n!$ permutations based on the ordinal ranking of $\vec{s}_i$. For example, $\vec{s}_0 = [x_0, x_1, x_2] \rightarrow \pi_1$ because $\vec{s}_0$ has three increasing values as $x_0>x_1>x_2$, which fits the form of $\pi_1$ as shown on the left side of Fig.~\ref{fig:timeseries_to_ordinal_partition_network}-(c). We continue this categorization process until the end of the time series. With the permutation sequence known, we generate an ordinal partition network $G$ by setting all $n!$ possible permutation types as nodes and forming edges $E$ between nodes $\pi_i$ and $\pi_j$ when a transition from $\pi_i$ to $\pi_j$ occurs. As an example, in Fig.~\ref{fig:timeseries_to_ordinal_partition_network}-(d) the first edge is formed between nodes $\pi_1$ and $\pi_2$ as those were the first two permutations from the time series. Continuing this process creates the square network in Fig.~\ref{fig:timeseries_to_ordinal_partition_network}-(d) through the permutation sequence $\pi_1 \rightarrow \pi_2 \rightarrow \pi_6 \rightarrow \pi_6 \rightarrow \pi_6 \rightarrow \pi_3 \rightarrow \pi_1$. 

\begin{figure}[h] 
    \centering
    \includegraphics[scale = 0.625]{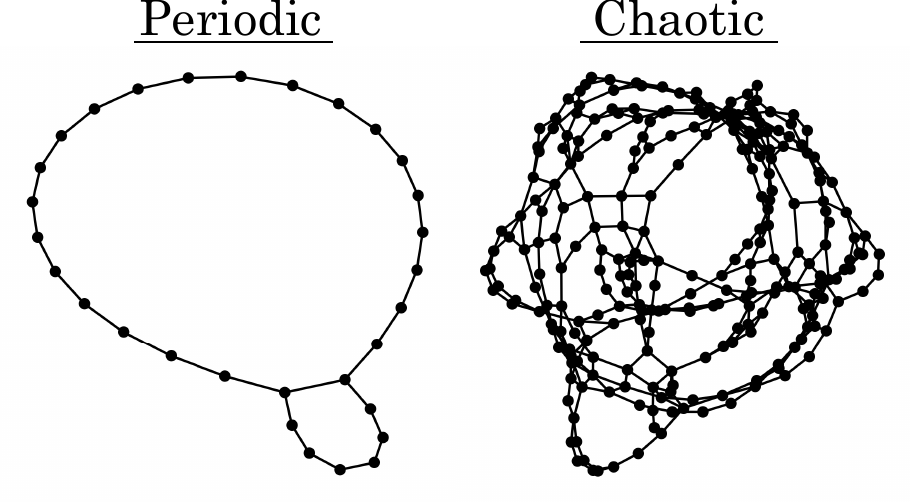}
    \caption{Examples of ordinal partition networks of a (left) periodic and (right) chaotic time series with permutation dimension $n=6$.}
    \label{fig:periodic_v_chaotic_ordinal_partition_network}
\end{figure}

As a general observation of ordinal partition networks, chaotic time series tend to form  complicated and seemingly disorganized networks while periodic time series generate relatively simple structures with few loops. Figure~\ref{fig:periodic_v_chaotic_ordinal_partition_network} demonstartes the complexity difference between an ordinal partition network generated from a peridoic and chaotic time series.
In the work done by McCollough et al.~\cite{McCullough2015}, common network summary statistics such as mean-out-degree and the number of nodes were used to demonstrate how  ordinal partition networks function to capture the complexity of time series. However, it was not until our previous work in~\cite{Myers2019} that a test for dynamic state detection based on persistent homology, a tool from topological data analysis, was developed. This method will be discussed in Section~\ref{sec:persistent_entropy_of_network}.

\section{Persistent Homology}
Persistent Homology is a flagship tool of Topological Data Analysis (TDA), an emergent field in computational topology. The motivation for this analysis technique is that data sets have an inherent shape in some vector space, and this shape can have meaning. This section gives an informal introduction to persistent homology. Specifically, we will introduce both sub level set persistence and persistent homology of complex networks. For precise definitions and formulations, the reader is directed  to~\cite{Munkres1993, Munch2017, Cohen-Steiner2006, Ghrist2014, Edelsbrunner2013, Edelsbrunner2008}. 

\subsection{Sub level set persistence} \label{sec:sublevel}
A standard approach to applying persistent homology to the $p$-$q$ projections would be to use a Rips complex filtration by expanding balls of radius $\epsilon$ about each point in the $p$-$q$ space and computing the homology of the resulting simplicial complex at each radius $\epsilon$. 
The hypothesis would then follow that the annular point cloud projections would be readily identifiable via the persistence of a one dimensional hole in the persistence diagram. 
However, we note that often times periodic time series data produces a $p$-$q$ projection which is bounded and regular, where some points populate the center.

\begin{figure}[h]
	\centering
	\includegraphics[width=.45\textwidth]{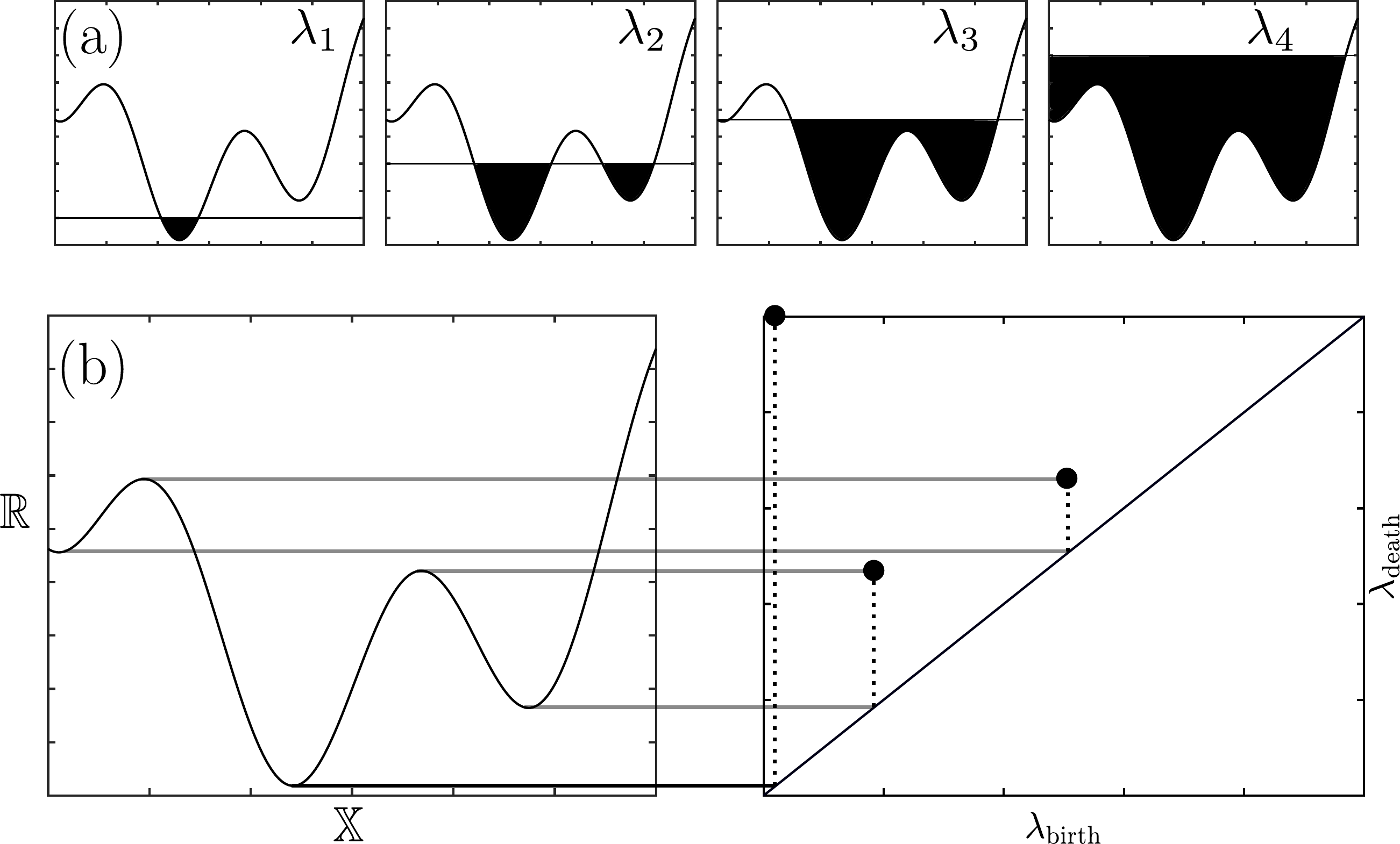}
	\caption{The rising level set $\lambda$ one  a one dimensional function with filled in regions indicating the collection of $L_{\lambda} = \{x:f(x)\leq\lambda\} = f^{-1}([-\infty,\lambda])$ and (bottom) the corresponding persistence diagram showing for how long each homology class persists. In this case, only class $H_0$ is an option. }
	\label{fig:0D_sub}
\end{figure}

In our framework, we take the kernel density estimate of the $p$-$q$ projection as a real valued height function $f:{X}\to {R}$ (see section~\ref{sec:PS Test}). On the function $f$, define $\lambda$ sub-level sets such that 
\begin{equation}
L_{\lambda} = \{x:f(x)\leq\lambda\} = f^{-1}([-\infty,\lambda])
\end{equation}
where the condition $L_{\lambda_1}\supseteq L_{\lambda_2}$ must hold for any pair $\lambda_1>\lambda_2$. The collection of the sets $\{L_\lambda\}_{\lambda\in {R}}$ generates a filtration with the level set being the index set. We give Fig.~\ref{fig:0D_sub} to elucidate this concept. Consider this one-dimensional function mapping $X$ to $R$ with local minima and maxima. The level set $\lambda$ can be most intuitively imagined as a water level which rises through the function during the filtration. When a minima is reached, a new region begins to ``fill in" and thus a new 0 dimensional homology generator is ``born". When the water level rises to a point where multiple regions merge, one of the connected components ``dies" as it merges with the other. 

\begin{figure}[t]
	\centering
	\includegraphics[width=.45\textwidth]{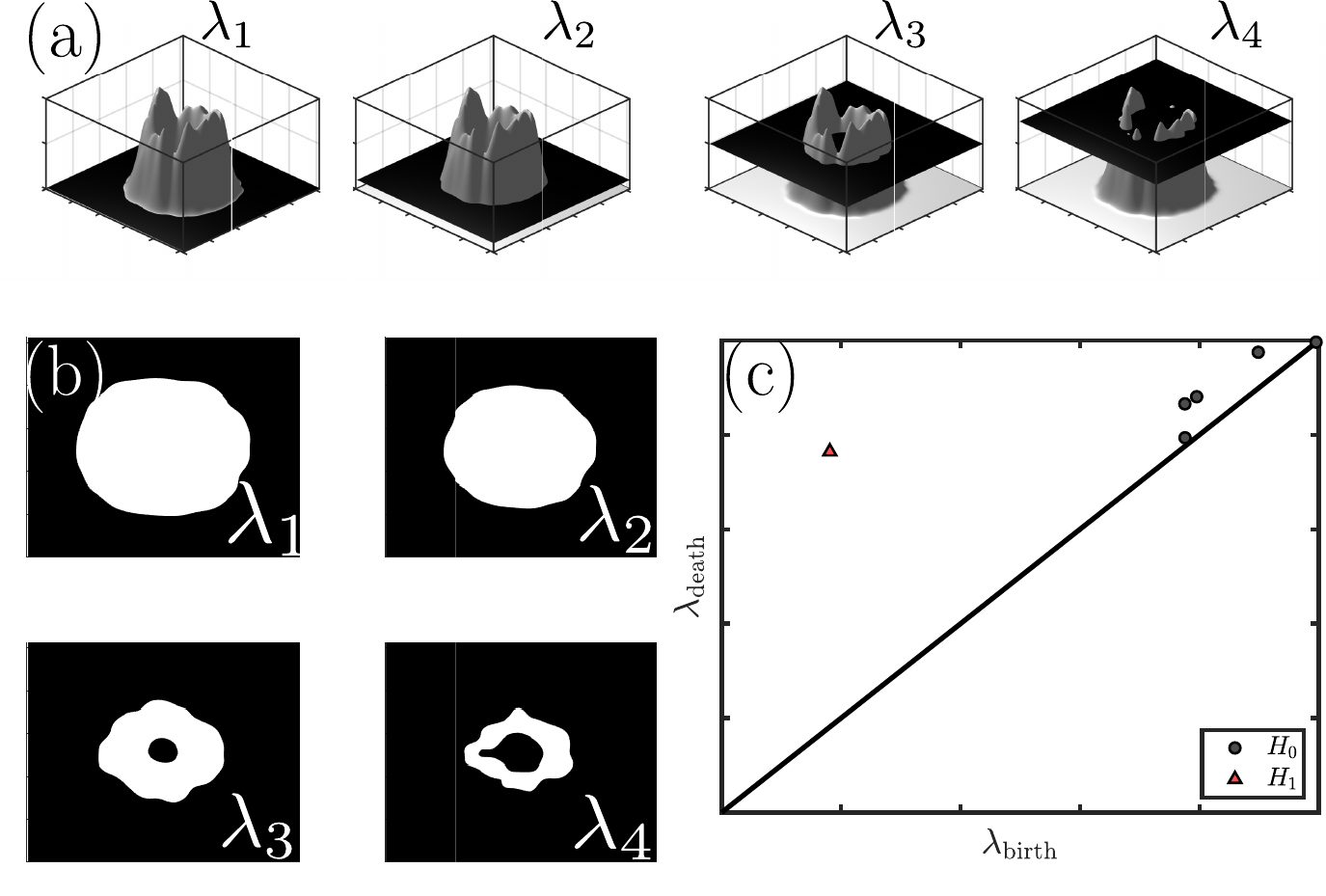}
\caption{(a) The rising level set $\lambda$ on the two dimensional function $f:X\to R$. (b) The sets of $x$ suction that $\{x:f(x)\leq\lambda\} = f^{-1}([-\infty,\lambda])$ and (c) the corresponding persistence diagram.  }
\label{fig:1D_sub}
\end{figure}

This analogy can be extended to two-dimensional height functions as shown in Fig.~\ref{fig:1D_sub}. As before, the set $\lambda$ rises through the function and tracks for how long features persist. However, instead of tracking the number of filled in regions like in Fig.~\ref{fig:0D_sub}, more intricate topological features are computed at each cross section (Fig.~\ref{fig:1D_sub}(b)). The 0 dimensional homology generators $H_0$ correspond to connected components while the 1 dimensional homology generators $H_1$ correspond to loops in the topology. The final result is the persistence diagram (Fig.~\ref{fig:0D_sub}(c)). The persistence diagram is noted with ${\mathsf{D}} = \{
(r_j,b_j,d_j):j = 1,\dots,|D|\}$ where the terms $r_j$, $b_j$, and $d_j$
are the homology order, birth level, and death level, respectively. 

\subsection{Persistent Homology of Networks}  \label{sec:persistent_entropy_of_network}
\begin{figure}[h] 
    \centering
    \includegraphics[scale = 0.625]{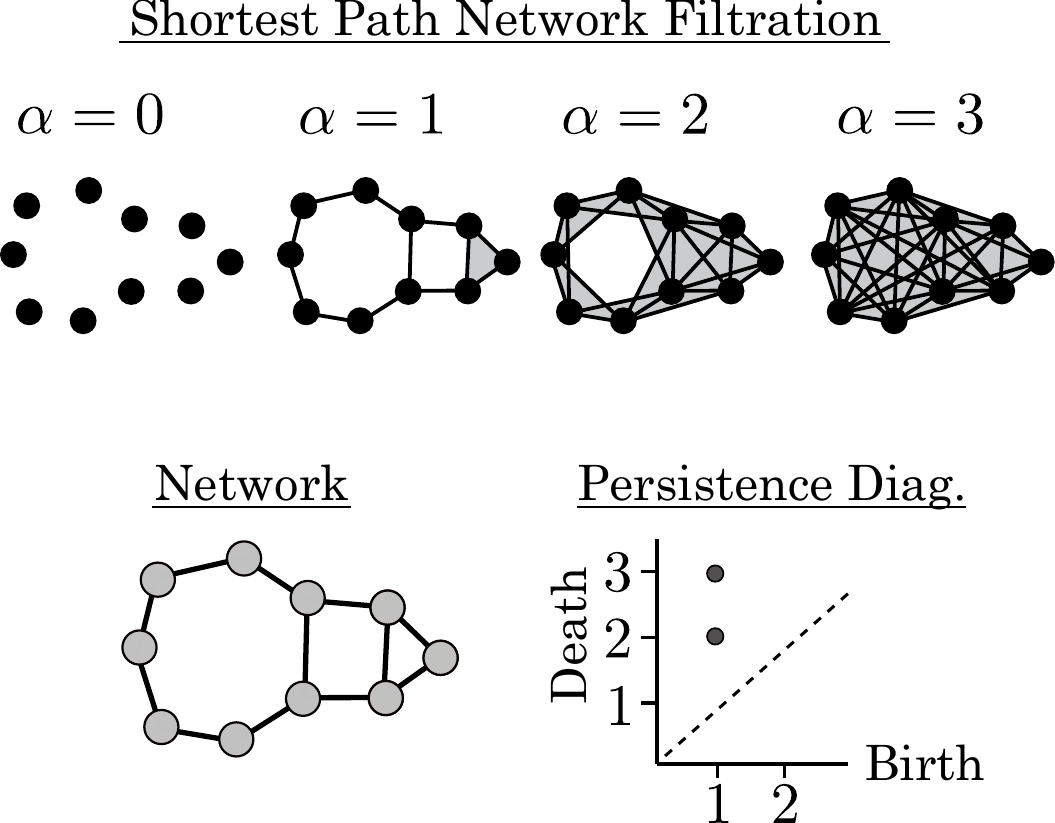}
    \caption{Example application of persistent homology applied to an unweighted and undirected network.}
    \label{fig:pers_diag_from_network}
\end{figure}

We will now introduce persistent homology of undirected and unweighted networks as a tool for measuring the underlying shape of a network. Persistent homology is used to measure the significance of certain homology groups such as loops and voids by applying a distance filtration $\alpha$ through either the Vietoris–Rips or Cech complex. This method of analysis works well for summarizing the structure and features of an undirected network.
As an example of how persistent homology can be applied to networks, consider the set of vertices and edges shown in the network on the bottom left of Fig.~\ref{fig:pers_diag_from_network}. This network is undirected and unweighted with the distance between directly connected vertices being $\rho = 1$, while the distance for in-directly connected vertices is the shortest path distance.
We now implement the $\rho = \alpha$ filtration until all components are connected. This starts with $\alpha = 0$, which returns the original vertices of the network as shown in the top left of Fig.~\ref{fig:pers_diag_from_network}. Next, at $\alpha = 1$, edges are formed between directly connected nodes, which creates a new face and two loops. We will track the birth of these loops through a persistence diagram as shown on the bottom right of Fig.~\ref{fig:pers_diag_from_network}. At $\alpha = 2$, the smaller loop fills in marking its death in the persistence diagram. Then, at $\alpha = 3$, The larger loop fills in signifying its death and ending the filtration. The final result of this filtration is the persistence diagram, which summarizes the significance of loops in the network through its (birth, death) coordinate.
For a more detailed overview of this method, please reference our original works in~\cite{Myers2019}.

\section{Detecting Chaos with Persistence} \label{sec: detecting_chaos_w_pers}
The two methods for detecting chaos with persistent homology are based on adaptations of the previously described chaos tests: the 0--1 test and ordinal partition networks. By implementing persistent homology, we are able to automate the process for detecting chaos and increase the noise robustness of the methods due to the stability of the persistence diagram~\cite{Cohen-Steiner2006}.

\subsection{PS Test} \label{sec:PS Test}
\begin{figure}[h]
	\centering
	\includegraphics[width=.425\textwidth]{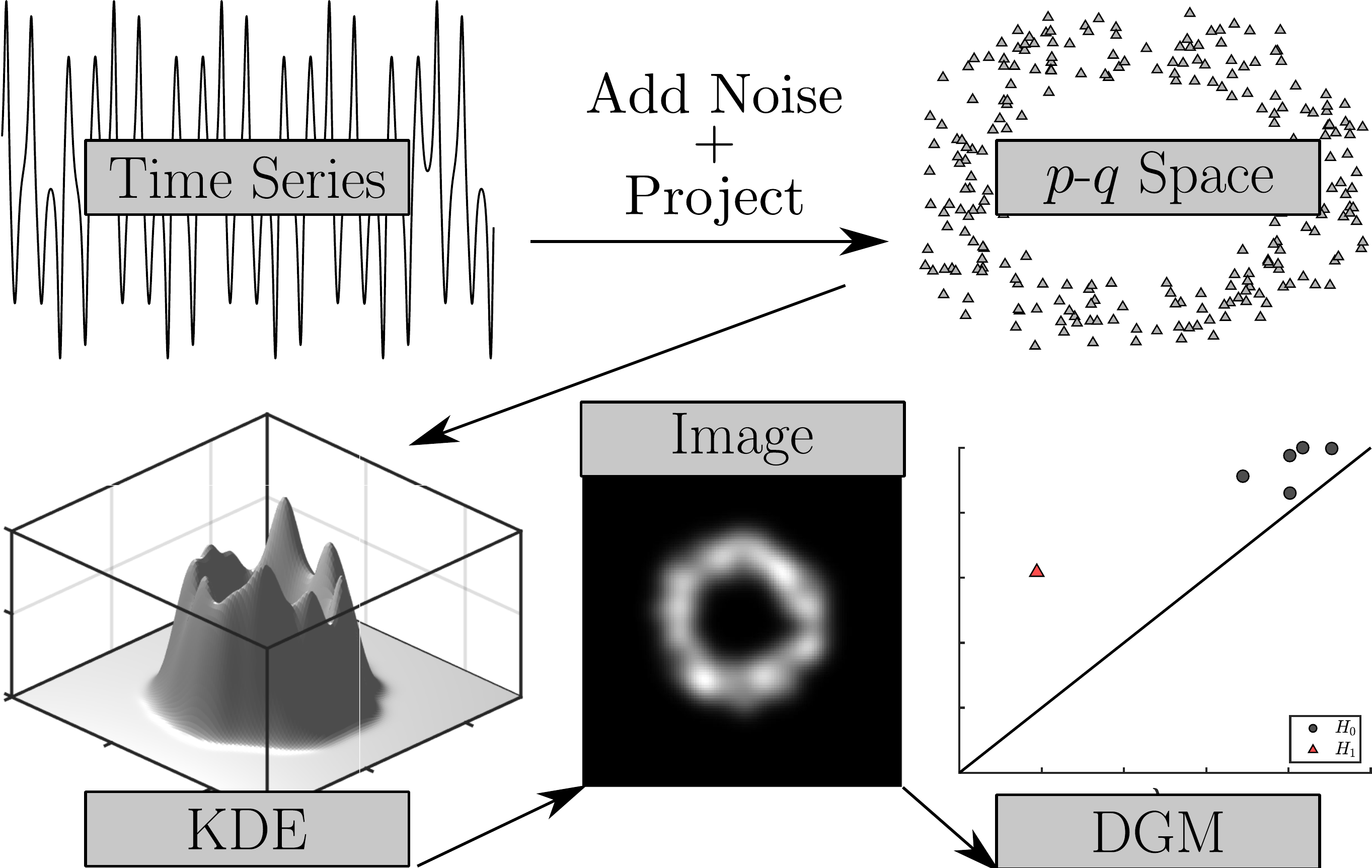}
	\caption{The pipeline for converting a time series into a persistence diagram using the KDE of a $p$-$q$ planar projection. }
	\label{fig:PS1_Pipeline}
\end{figure}

In~\cite{Tempelman2019}, we introduced a method for delivering a binary diagnostic for chaos in time series data based on persistent homology. The method utilizes the topological structures of $p$-$q$ projections and returns a Persistence Score $PS$ which summarizes the dynamics of the data. This section will only give a formulaic procedure regarding the application of the method; the methodological justifications can be found in~\cite{Tempelman2019}. 
The time series data is projected into $p$-$q$ space per Eq.~\eqref{eq:pq}. Periodic time series give bounded $p$-$q$ projections with points most heavily populating an annular region about the origin, however it is sometimes the case that points will populate the center regions as well. To pull meaningful topological information form these point clouds, we first take the kernel density estimate (KDE) using the diffusion method presented in~\cite{Botev2010}. Next, a Gaussian smoothing filter is applied to the height function per
\begin{equation}
G_h = \frac{1}{2\pi h^2}{\rm e}^{-(x^2+y^2)/2h^2}
\end{equation}
with a kernel bandwidth of $h=1.3$. The KDEs are then converted into a gray-scale image by assigning a pixel intensity value in proportion to the function value; these values are normalized to have a peak value of 1. The sub-level set persistence scheme is then be applied to the image data via the open source software DIPHA\footnote{https://github.com/DIPHA/dipha} to yield the persistence diagram. This pipeline is graphically displayed in Fig.~\ref{fig:PS1_Pipeline}. Only the 0D homology generators are considered. Persistence points which have a birth value of $\lambda<0.01$ are discarded as well since these can be thought of as equivalent to topological noise. 

A unique persistence diagram is generated for each selection of $c\in(0,\pi)$. Typically, a periodic time series will yield a persistence diagram with birth and death times near 1 so that the distance to the origin is near $\sqrt{2}$ as shown in~\cite{Tempelman2019}. 
The chaotic time series typically produce persistence points near the center of the upper left triangle. Although their expected value of random topological fields in persistence space is difficult to resolve~\cite{Tempelman2019,Adler2010,Adler2014}, previous research empirically suggests that the expected value of a random topological field falls near the center of the upper left triangle~\cite{Adler2010}. Thus, the diffuse $p$-$q$ projections should yield persistence data which is typically much closer to the origin than $\sqrt{2}$, and this is indeed the case~\cite{Tempelman2019}. Thus, the $PS$ score is taken as the mean distance of the 0D persistence points to the origin of the persistence diagram. To enhance reproducibility, the ensemble mean of 200 persistence diagrams generated from 200 unique selections of $c$ used. This is computed via 
\begin{equation}
PS =\Bigg \langle \sum_{j=1}^{N_i} \left\{\frac{\ \sqrt{(d_j^2 + b_j^2)}}{N_i}, (b_j,d_j)\in {\mathsf D}_i\right\} \Bigg \rangle, 
\label{eq:PS1}
\end{equation}
where $PS\in[0,\sqrt{2}]$, ${\mathsf D}_i$ is the $i$th persistence diagram, and $\langle \ \ \rangle$ notes the ensemble mean. If $PS\approx\sqrt{2}$, a periodic diagnostic is returned while a $PS\approx \sqrt{2}/2$ indicates a chaotic diagnostic.

\subsection{Persistent Entropy of Ordinal Partition Networks}\label{sec: persistence_on_networks}
Returning to the ordinal partition networks and their persistent homology, our next goal is to analyze the resulting persistence diagrams $P$ through the lifetimes of the features. The lifetime of a feature is defined as $L_i = D_i - B_i$, where $B_i$ and $D_i$ are the birth and death values of the $i^{th}$ feature in the persistence diagram, which is stored as the point $(B_i, D_i)$. We can now use the entropy of the persistence diagram (persistent entropy) $E(P)$ as a summary statistic to analyze the dynamic state as a complexity measurement.

Persistent entropy was first developed by Chintakunta et al.~\cite{Chintakunta2015}. The statistic uses the entropy of the lifetimes from a persistence diagram, where entropy is calculated as information entropy. Specifically, this summary statistic is defined as
\begin{equation}
E(P) = - \sum_{x \in P} \frac{{\rm pers}(x)}{\mathscr{L}(P)}\log_2\left(\frac{{\rm pers}(x)}{\mathscr{L}(P)}\right),
\label{eq:pers_ent}
\end{equation}
where $\mathscr{L}(P) = \sum_{x \in P} {\rm pers}(x)$ is the sum of lifetimes of points in the diagram with ${\rm pers}(x) = L_i$.
We normalize $E(p)$ as $E'(P) = {E(P)}/{\log_2\big(\mathscr{L}(P))}$. In our previous work~\cite{Myers2019} we showed that the normalized persistent entropy $E'(P)$ was able to accurately detect chaotic in comparison to periodic dynamics through a threshold of 0.8, with  $E'(P) \geq 0.8$ signifying chaos.

\section{Noise Models} \label{sec: Noise}

The traditional uncorrelated Gaussian noise model (White noise) can be constructed as a noise vector $\xi(t)$ for which the relation holds $\langle \xi(t),\xi(s) \rangle =2D \delta(t-s)$, where $\delta$ is the Kronecker delta function and $D$ defined as the noise intensity.  This noise model is uniformly broadband is the de-facto standard for artificially construing noise.  
However, as mentioned in the introduction, correlated noise processes are often of interest. The correlated noise models are typically referred to as colored noise with the most common being red, pink, white, blue, and purple.
In this paper, we construct the colored noise models with the help of the digital signal processing toolbox available in \textsc{matlab} which follows the noise model formulation of~\cite{Kasdin1995}. 

The colored noise models are most easily described by their spectral characteristics. 
Pink noise and Brown noise decrease in power at a rate of 3dB per octave and 6 dB per octave, respectively. 
Alternatively, blue noise and violet noise increase in spectral power at 3 dB per octave and 6 dB per octave, respectively~\cite{Kasdin1995}.  These noise models produce a power spectral density (PSD) of the form $S(f)\sim 1/f^\alpha$ where $\alpha = -2,-1,0,1,2$ yield purple, blue, white, pink, and red noise respectively. 


\begin{figure}[h]
	\centering
	\includegraphics[width=0.45\textwidth]{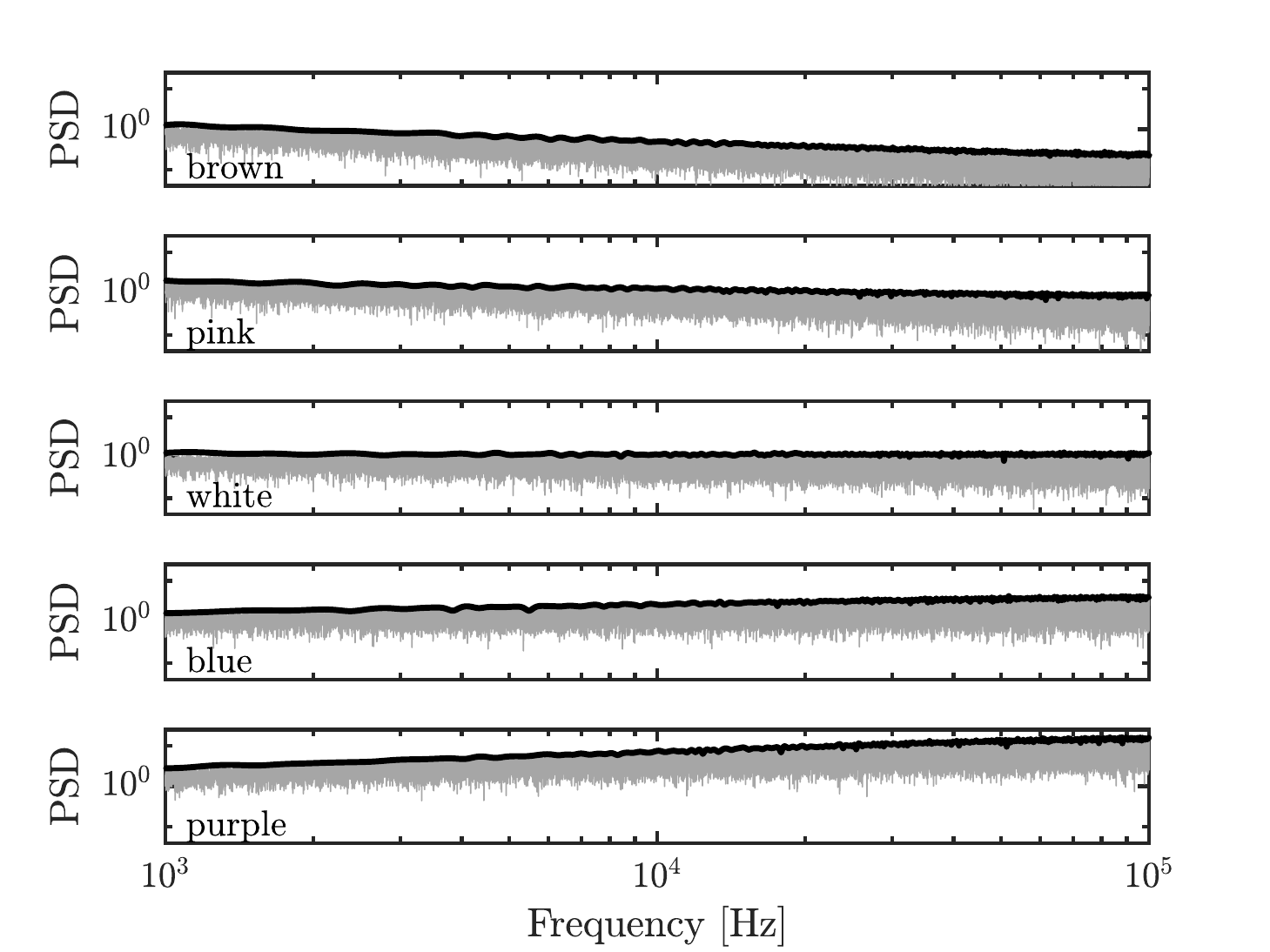}
	\caption{Power spectral densities of five different noise colors (red, pink, white, blue, and violet) normalized so that the PSDs are approximately equal in power at a frequency of 100 Hz. }
\end{figure}
We use the signal to noise ratio (SNR) between the simulated time series and the noise vector as the measure for noise intensity in this study. 
This is the most common tool for measuring noise levels in time series. The SNR is given in dB as
$
\text{SNR} = 20\log_{10}\left(
{\text{rms}_{\text{signal}}}/{
	\text{rms}_{\text{noise}}
}
\right)$,
where an SNR of 30 dB is considered a moderate noise level and an SNR of 15 dB is typically set as a minimum threshold to extract meaningful information from a signal.

We use a desired SNR level to scale the synthetically generated colored noise vectors.  
For each signal, the root mean square of the uncontaminated value is measured. With this, a colored noise vector is scaled appropriately to meet the desired SNR value and added to the original time series to construct the test data $\tilde x = x + \epsilon\xi$, where $\tilde x$ is the noisy data, $x$ is the original signal, $\xi$ is a vector of noise, and $\epsilon$ is the scaling factor.

\section{Results} \label{sec: Results}
\begin{figure}[b!] 
	\centering
	\includegraphics[scale = 0.29]{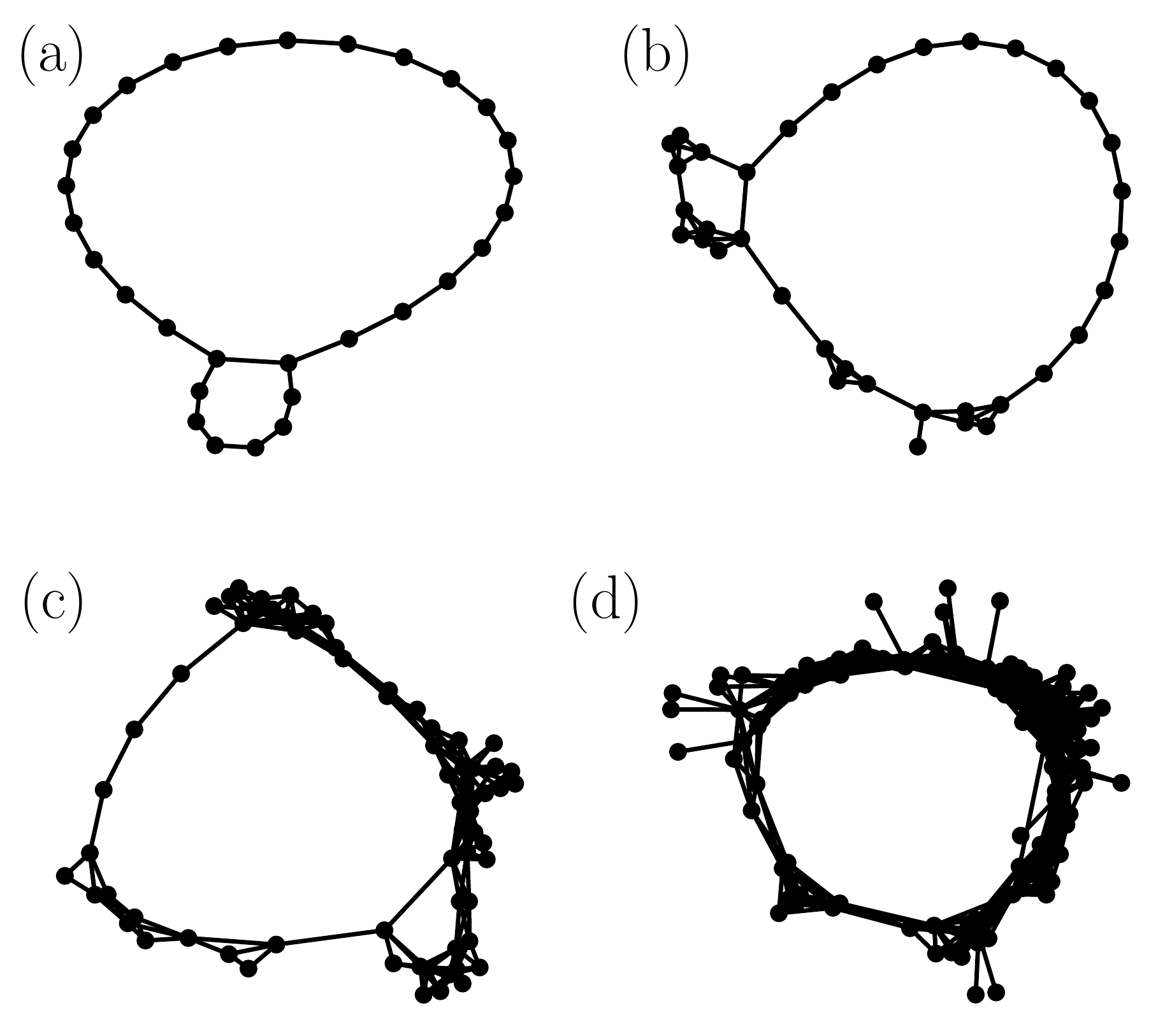}
	\caption{Sample ordinal partition networks of dimension $n=6$ from the periodic Rossler system ($a = 0.25$) with increasing white noise levels. In Fig.~(a),~(b),~(c),~and~(d) the additive white noise SNR is $\infty, 40, 30$, and $20$, respectively.}
	\label{fig:rossler_networks_SNR_inf_40_30_20}
\end{figure}

\begin{figure*}[h]
	\centering
	\includegraphics[width=0.92\textwidth]{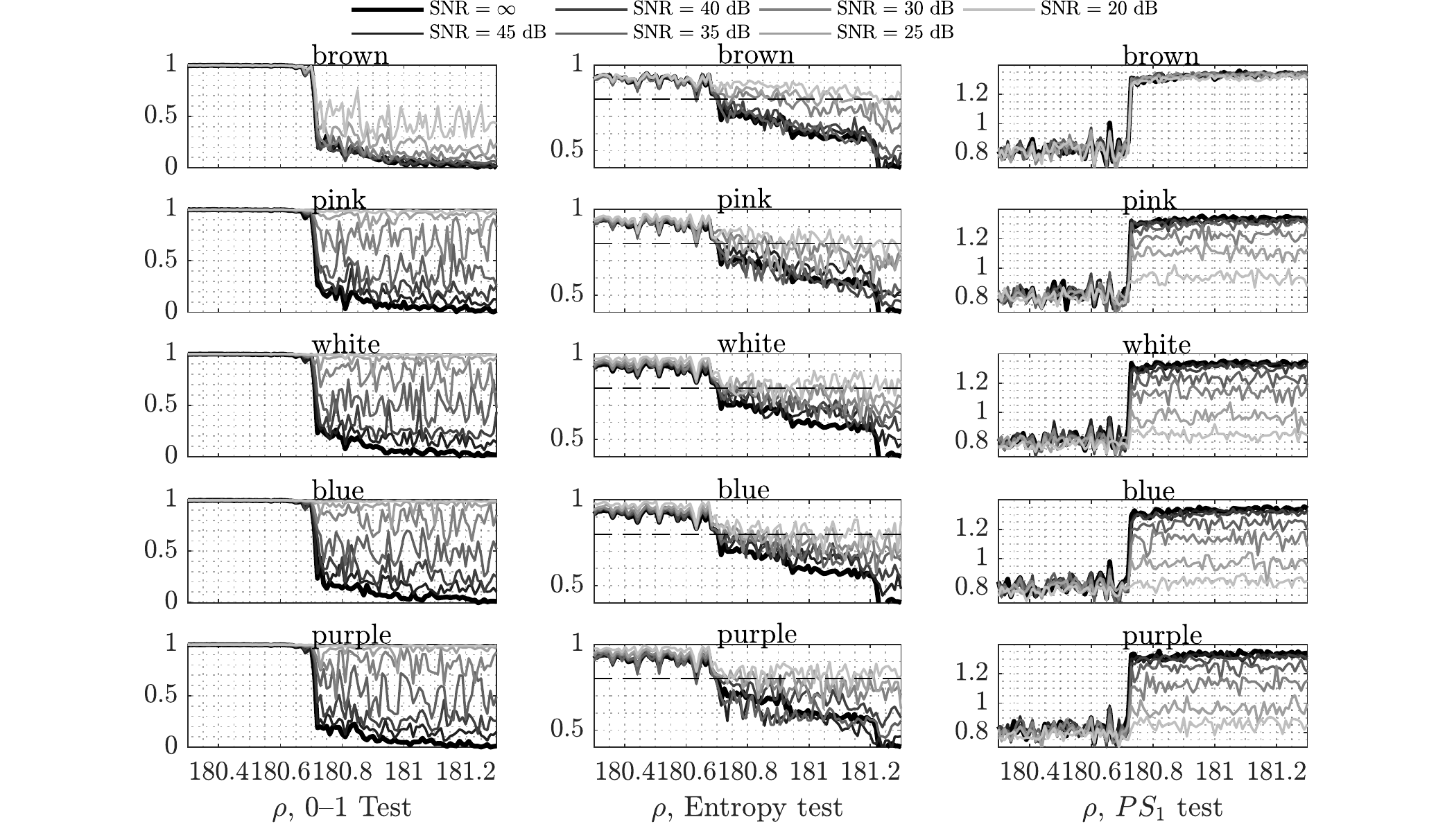}
	\caption{Effects of colored noise on the 0--1 test and $PS_1$ scores for the Lorenz equations.  }
	\label{fig:results_lorenz}
\end{figure*}
To examine the effects of the colored noise on the state detection tools, numerical studies on the Lorenz and Rossler models are carried out. The Lorenz and Rossler equations are classic dynamical systems that are known to exhibit a rich variety of dynamics over narrow spans of bifurcation parameters~~\cite{Letellier1995,Lorenz1963}. The Lorenz (Eq.~\eqref{eq: Lorenz}) and Rossler (Eq.~\eqref{eq: Rossler}) models are given as
\begin{equation} \label{eq: Lorenz}
	\dot{x} = \sigma(y-x), \ \ \ \ \ \
	\dot{y} = x(\rho-z)-y, \ \ \ \ \ \
	\dot{z} = xy-\beta{z}, 
	\end{equation}
\begin{equation} \label{eq: Rossler}
    \dot{x} = -y-z, \ \ \ \ \ \
 	\dot{y} = x+ay, \ \ \ \ \ \
    \dot{z} = b + z(x-c),
\end{equation}
where, for the Lorenz equation, $\beta = 8/3$, $\sigma = 10$, and $\rho\in[180.3,\ 181.3]$, and for the Rossler equation $a\in[0.25, \ 0.55]$, $b=2$, and $c = 4$.

\begin{figure*}[h]
	\centering
	\includegraphics[width=0.92\textwidth]{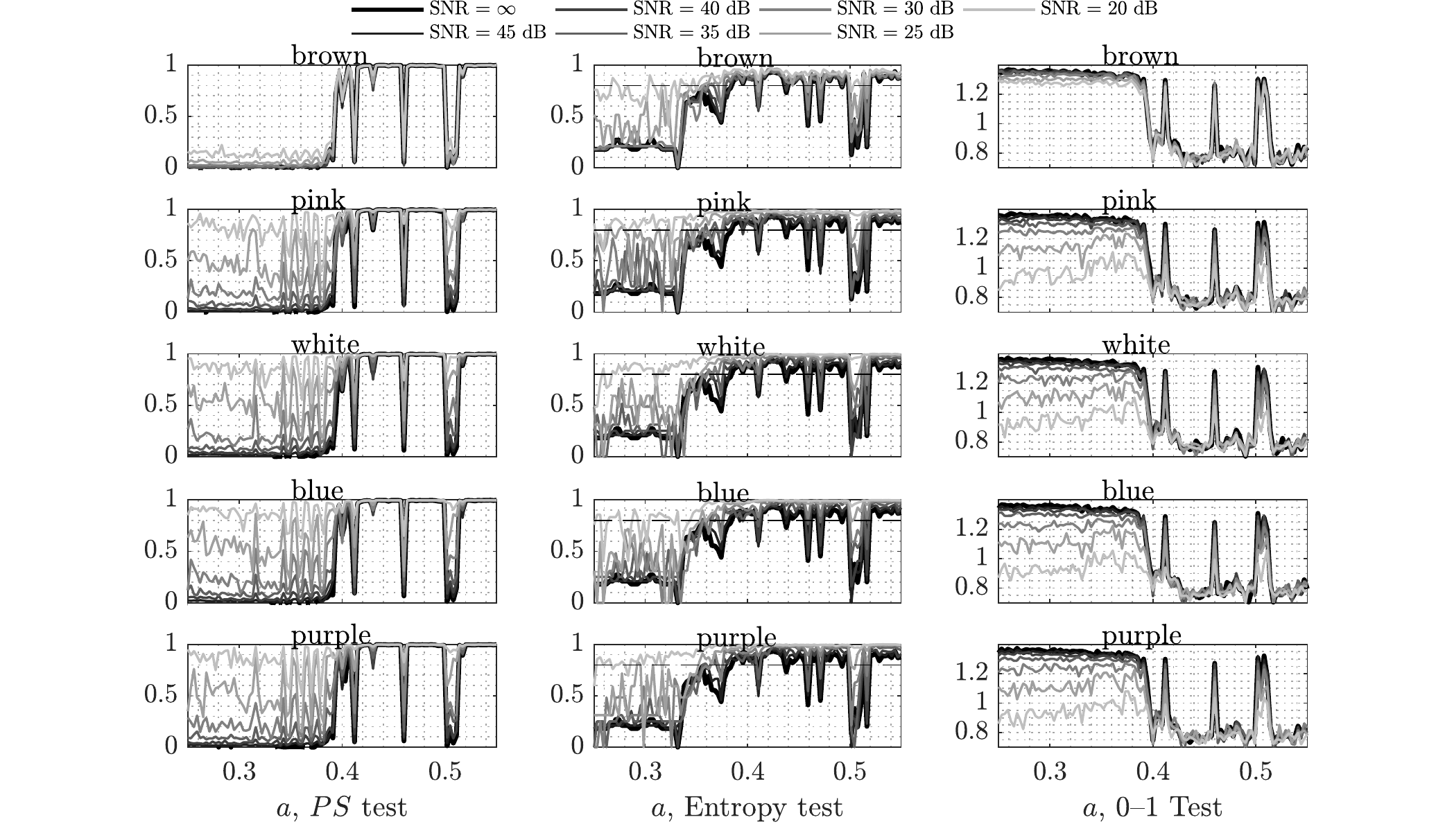}
	\caption{Effects of colored noise on the 0--1 test and $PS_1$ scores for the Rossler equations.  }
	\label{fig:results_rossler}
\end{figure*}

The effects of the additive noise on the respective diagnostic tools are given in Figs.~\ref{fig:results_lorenz} and~\ref{fig:results_rossler} for the Lorenz and Rossler models, respectively. The time series data is acquired for both models via the \textsc{matlab} ODE45 routine. The time steps are set to $\Delta t = 0.001$ and $\Delta t = 0.01$ seconds of the Lorenz and Rossler models, respectively. For both models, a transient cut-off of 100 seconds is used. 
A baseline periodicity score is given in Figs.~\ref{fig:results_lorenz} and~\ref{fig:results_rossler} as the bold black line. The scores which encompasses the dark lines are taken from time series data with no noise and were shown in~\cite{Tempelman2019} to agree with the long time limit convergence of the accepted 0--1 test. For ordinal partition networks, a permutation dimension of $n=6$ was used for both systems and permutation delays of $\tau = 108$ and $\tau = 170$ for the Lorenz and Rossler systems, respectively.


Fig.~\ref{fig:results_lorenz} shows that for the Lorenz model, the effects of noises are heightened as $\alpha$ increases. For the case of $\alpha = -2$ (red noise), there is seemingly little to no effect on the $PS_1$ scores even for an SNR of 20 dB. As the value of $\alpha$ increases, the effects of noise are much more obvious. There is a falling-of from the accepted noise-free scores when SNR values hit 30 dB for blue noise, and this fall off is observable at an SNR of 35 dB for white, pink, and red noises. Note that the $PS_1$ test remains more stable than the 0--1 correlation test for all $\alpha$ whit false positives appears even for $\alpha = -2$. The entropy based scores are similarly impacted by the noise, however the $\alpha = -2$ noise yielding false positives at an SNR of 30 to 25 dB. 

Interestingly, the same effects are not replicated for the Rossler system given in Fig.~\ref{fig:results_rossler}. It is again observed that the $PS_1$ scores remain fixed to their noise-free diagnostic for all tested SNR values when $\alpha = -2$. However, the false positives seen in the 0--1 test are not observed on the Rossler model meaning that the lower frequency oscillations of red noise are more detrimental to high-frequency time series (the fundamental frequency of the Lorenz model is much greater than that of the Rossler model for the parameter values selected for this study). It shown in Fig.~\ref{fig:results_rossler} that for $\alpha >-1$, the effects of noise become obvious for an SNR of 30 dB for the $PS_1$ scores and 0--1 correlation scores indicating. Furthermore, the effects of noise color are less prominent for the Rossler model, however it is still observed that the effect of the noise is heightened as $\alpha$ increases. 

To better understand the effects of noise on ordinal partition networks, we generated four networks with increasing levels of noise as shown in Fig.~\ref{fig:rossler_networks_SNR_inf_40_30_20}. Specifically in Fig.~(a),~(b),~(c),~and~(d) the additive white noise SNR is $\infty, 40, 30$, and $20$, respectively. These networks were from the periodic Rossler system at $a=0.25$. In Fig.~(a), (b), and (c) the overall structure of the ordinal partition network prevails through the additive white noise with persistent entropy values of 0.21, 0.24, and 0.25, respectively. However, at an SNR of 20 dB in Fig.~(d) the network loses the majority of its structure. As a general result, this shows that persistent entropy of ordinal partition networks should withstand up to approximately 30 dB of additive noise before breaking down. To accommodate for higher levels of noise, we propose implementing an algorithm to assign weights to the network edges based on the number of transitions.


\section{Conclusion} \label{sec: Conclusion}

Previous work on the effects of noise for state detection methods principally focus on Gaussian uncorrelated noise.
Because colored noise can often times present itself in mechanical and electrical systems~\cite{Schueller2006,Ding2013,kasdin1995discrete}, the influence  of noise correlation needs to be studied on dynamic state detection, or chaos detection methods. 
In this work, we have examined the performance of several dynamic state detection methods when faced with correlated $1/f^{\alpha}$ noises. 
Specifically, the 0--1 test, $PS$ scores, and normalized entropy tests are studied.

It is shown through simulations of the Lorenz and Rossler models that the effects of noise tend to be more adverse on these chaos detection tools when the value of $\alpha$ increases. Moreover, the $PS_1$ scores remain more stable across all $\alpha$ than the 0--1 test. The normalized entropy scores are non-binary and therefore difficult to directly compare to the $PS_1$ scores or 0--1 scores, although it is observed that the normalized entropy method delivers accurate results according to its 0.8 threshold for SNR values up to 30 dB even for high frequency ($\alpha = 2$) noise correlations. In future work, we believe that the noise robustness of the ordinal partition networks can be increased by including edge weights.

This study offers a preliminary glance at the impact of noise correlation on dynamic state detection methods.
The empirical results of this study indicate that raising the value of $\alpha$ in $1/\alpha$ correlated noise processes makes the detection of chaotic dynamics more difficult. While this empirical study is insightful, a more formal study should be pursued to theoretically validate the empirical observations that chaos detection methods become less effective when $\alpha$ increases for $1/f^\alpha$ correlated noise processes.


\section*{Acknowledgments}
This material is based upon work supported by the National Science Foundation under grant
nos. CMMI-1759823 and DMS-1759824 with PI FAK.

\bibliographystyle{plain}
\bibliography{IDETC2020}

\end{document}